\documentclass[10pt,journal,compsoc]{IEEEtran}
\ifCLASSINFOpdf
\else
\fi
\usepackage[space,nocompress]{cite}
\usepackage{graphicx,caption,amssymb,amsmath,array}
\usepackage{color,float}
\usepackage{subfigure,epsfig}
\usepackage{ragged2e}
\usepackage{enumitem}
\usepackage{footnote}

\usepackage{algorithm}
\usepackage{algorithmic}
\usepackage{multirow}
\usepackage{float}
\usepackage{url}
\usepackage{enumitem}
\usepackage{array}
\usepackage{booktabs}
\usepackage{bm}
\usepackage{color}

\setlist[itemize]{leftmargin=*}
\setlist[enumerate]{leftmargin=*}
\begin{document}
\title{Hierarchical User Intent Graph Network for Multimedia Recommendation}


% \author{\IEEEauthorblockN{Yinwei\IEEEauthorrefmark{1},
% Xiang Wang\IEEEauthorrefmark{1},
% Xiangnan He\IEEEauthorrefmark{2},~\IEEEmembership{Member,~IEEE}, \\
% Liqiang Nie\IEEEauthorrefmark{3},~\IEEEmembership{Member,~IEEE}, 
% and Tat-Seng Chua\IEEEauthorrefmark{1},~\IEEEmembership{Member,~IEEE}}
% \IEEEauthorblockA{\IEEEauthorrefmark{1}School of Computing, National University of Singapore, Singapore.}
% \IEEEauthorblockA{\IEEEauthorrefmark{2}, University of Science and Technology of China, Hefei, China}
% \IEEEauthorblockA{\IEEEauthorrefmark{3}College of Computer Science and Technology, Shandong University, Qingdao, Shandong, China.}
% \thanks{Xiang Wang and Liqiang Nie are the corresponding authors.}}
\author{Yinwei~Wei~\IEEEmembership{, Member, IEEE},
        Xiang~Wang~\IEEEmembership{, Member, IEEE},
        Xiangnan He~\IEEEmembership{, Member, IEEE},\\
        Liqiang~Nie~\IEEEmembership{, Senior Member, IEEE},
        Yong~Rui~\IEEEmembership{, Fellow, IEEE},
        and Tat-Seng Chua~\IEEEmembership{, Member, IEEE}% <-this % stops a space
\IEEEcompsocitemizethanks{
\IEEEcompsocthanksitem {Y. Wei, X. Wang, and T. S. Chua are with the School of Computing, National University of Singapore, Singapore. \protect(e-mail: weiyinwei@hotmail.com; xiangwang@u.nus.edu; chuats@comp.nus.edu.sg).}
\IEEEcompsocthanksitem {L. Nie is with the College of Computer Science and Technology, Shandong University, Qingdao, Shandong, China. \protect(e-mail: nieliqiang@gmail.com).}
\IEEEcompsocthanksitem {X. He is with the University of Science and Technology of China, Hefei, China. \protect(e-mail: xiangnanhe@gmail.com).}
\IEEEcompsocthanksitem{Y. Rui is with the Lenovo Group, Beijing, China. \protect(e-mail: yongrui@lenovo.com).}
}
}

\markboth{Journal of \LaTeX\ Class Files,~Vol.~14, No.~8, August~2015}%
{Shell \MakeLowercase{\textit{et al.}}: Bare Demo of IEEEtran.cls for IEEE Transactions on Magnetics Journals}



% for Transactions on Magnetics papers, we must declare the abstract and
% index terms PRIOR to the title within the \IEEEtitleabstractindextext
% IEEEtran command as these need to go into the title area created by
\IEEEtitleabstractindextext{%
\justifying  
\begin{abstract}
Understanding user preference on item context is the key to acquire a high-quality multimedia recommendation.
Typically, the pre-existing features of items are derived from pre-trained models (\textit{e.g.,} visual features of micro-videos extracted from some neural networks), and then introduced into the recommendation framework (\textit{e.g.,} collaborative filtering) to capture user preference.
However, we argue that such a paradigm is insufficient to output satisfactory user representations, which hardly profile personal interests well.
The key reason is that present works largely leave user intents untouched, then failing to encode such informative representation of users. 
% then failing to encode such finer-grained preferences of users.

In this work, we aim to learn multi-level user intents from the co-interacted patterns of items, so as to obtain high-quality representations of users and items and further enhance the recommendation performance.
Towards this end, we develop a novel framework, \emph{Hierarchical User Intent Graph Network}, which exhibits user intents in a hierarchical graph structure, from the fine-grained to coarse-grained intents.
In particular, we get the multi-level user intents by recursively performing two operations: 
1) intra-level aggregation, which distills the signal pertinent to user intents from co-interacted item graphs; 
and 2) inter-level aggregation, which constitutes the supernode in higher levels to model coarser-grained user intents via gathering the nodes' representations in the lower ones. 
% and summarizes the supernode's representations to capture coarser-grained user intents.
Then, we refine the user and item representations as a distribution over the discovered intents, instead of simple pre-existing features.
To demonstrate the effectiveness of our model, we conducted extensive experiments on three public datasets.
Our model achieves significant improvements over the state-of-the-art methods, including MMGCN and DisenGCN.
Furthermore, by visualizing the item representations, we provide the semantics of user intents.
\end{abstract}

}

\maketitle

\IEEEdisplaynontitleabstractindextext

% \fi
\IEEEpeerreviewmaketitle

\section{Introduction}
On multimedia content sharing platforms~(\textit{e.g.} Pinterest, Instagram, and TikTok)~\cite{NMCL}, users recently face the trouble of locating their interesting items from the overloaded information. 
Therefore, multimedia recommender systems have been designed as the core components of these service providers, which aims to discover the candidate items for users according to their tastes.

In order to suggest users a list of items meeting their individual interests, one of the key factors of the multimedia recommendation is to model the user preferences towards the content information. 
Present articles~\cite{VBPR,ACF,MMGCN} typically incorporate pre-existing features of items into the recommendation framework, especially the collaborative filtering~(CF), so as to profile the users based on historical interactions and item contents. 
Some representative methods, like VBPR~\cite{VBPR} and ACF~\cite{ACF}, tend to extract features of items via deep neural networks and then integrates them with the collaborative embeddings to learn the user preference. 
Despite their effectiveness, we argue that such a paradigm is insufficient to profile user preferences. 
One major reason is that these deep-learning-based feature extractors are tailored for computer vision~\cite{ResNet,DenseNet,SB1} and natural language processing~\cite{word2vec,BERT}, rather than the personalized recommendation. 
Hence, simply using such pre-existing features easily leads to suboptimal representations of user preference.

To solve this problem, one solution is to model the user intents to guide the user and item representation learning, which could distill the informative signal affecting the users' choices. 
In particular, the user intents are used to describe the motivations why people prefer some items. 
Based on the distribution over them, the item could be re-expressed according to users' motivation and the user preference reflects the motivation when she/he makes  decisions.
For instance, Liu~\textit{et al.}~\cite{UVCAN} devised a co-attention model to refine the items' features with the learned users' representations; Lei~\textit{et al.}~\cite{CDL} proposed the duel-net deep network to project the visual and user information into a latent semantic space to mimic the intents. 
However, such methods are limited to filter out the features no one concerned, instead of discovering the users' motivations. 
Although some models~\cite{REDAR,GraphRec} utilize the social information or user demographics to remedy this drawback, the gap between these external information and user intents still exists. 
\begin{figure}
	\centering
    \includegraphics[width=0.4\textwidth]{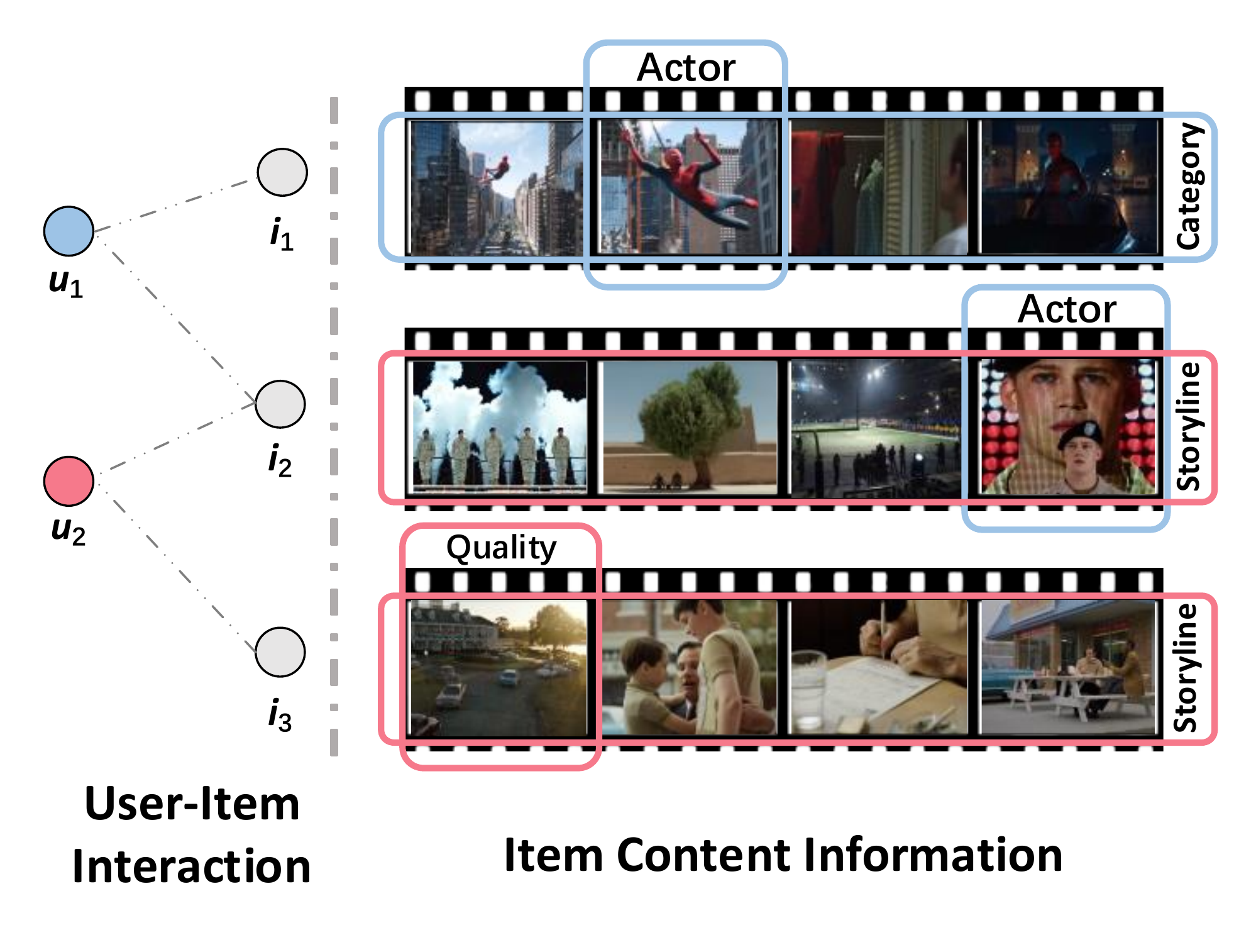}
    % \vspace{-5pt}
    \caption{Illustration of the users' multi-level intents on video content.}
	\label{fig:intro}
    \vspace{-20pt}
\end{figure}

Therefore, it is crucial to explicitly model the user intents and represent the users and items with the distributions on the learned intents. 
However, it is non-trivial due to the following challenges: 
\begin{itemize}
    \item The user intent cannot be acquired immediately, since it involves the information both from the users' behaviors and item content. Specifically, it is intuitively plausible that users' behaviors towards items reflect their intents; meanwhile, the content information tends to be the source of the user intents. 
    Considering these two types of information are heterogeneous, \textit{how to effectively organize them into a unified manner and distill the informative signal on user intents is the primary challenge in our task}.
    \item 
    The structure of user intents is not flat and each intent is not isolated in this structure. 
    As shown in Figure~\ref{fig:intro}, $u_1$ prefers the actor and category; whereas $u_2$ pays more attention to the visual quality and storylines of videos. 
    Whereinto, the visual quality and actor are shown by frames, while the storyline and category are summarized with the whole video. 
    It conveys the fact that user intents can be divided into different levels from the fine-grained (\textit{e.g. preferences for the frame)} to the coarse-grained~(\textit{e.g. interests for the whole video}). 
    Moreover, the coarser-grained ones can be summarized from the fine-grained ones. 
    Hence, \textit{how to model the hierarchical structure corresponding to users' multi-level intents is another unavoidable challenge we are facing}. 
\end{itemize}

Inspired by the recent progress in graph convolutional networks~(GCNs), we propose to employ the information aggregation mechanism to model a hierarchical user intent graph against the above challenges. 
Towards this end, we first construct a co-interacted item graph, where the edge corresponds to the item pair consumed by the same users, in order to integrate the information of item contents with that of user behaviors. 
Upon such a co-interacted graph, we then develop a Hierarchical User Intent Graph Network~(HUIGN), which learns the multi-level user intents' representations as well as the hierarchical structure. 
Specifically, we design two types of information aggregation: 1) intra-level aggregation which captures the local structure information~(\textit{i.e.} user co-interacted behaviors) for each node and distills the user intents via incorporating it with content information; and 2) inter-level aggregation which establishes some supernodes in the higher level and gathers information of nodes to summarize these supernodes as coarser-grained user intents. 
Iteratively performing these two aggregations, we explicitly learn the representation of multi-level user intents and the hierarchical structure from fine-grained intents to coarse-grained ones. 
Finally, we characterize the user and item as a distribution over the discovered intents to predict their interactions. 
To evaluate the proposed model, we conduct experiments on three publicly accessible datasets designed for the micro-video personalized recommendation. 
The results show that our proposed model achieves state-of-the-art performance. 
Besides, with the visualizations of items' representations, our model provides the answer to what the user intents are. 

In a nutshell, we summarize the main contributions as:
\begin{itemize}
    \item 
    We investigate the advantages and challenges of user intents learning in the multimedia recommendation and propose to explicitly model the multi-level user intents to optimize the user and item representations. 
    \item 
    We design a Hierarchical User Intents Graph Network to learn the user intents from the fine-grained to the coarse-grained. It incorporates the users' behavior with items' content information through the intra- and inter- level graph convolutional operations.
    \item 
    Conducting extensive experiments on three datasets, we demonstrate the effectiveness of our proposed method. 
    In addition, we provide the semantic representations of user intents through displaying the items' characters based on the user intents. 
    As a side contribution, we have released the data and codes to facilitate other researchers\footnote{https://github.com/weiyinwei/HUIGN.}.
\end{itemize}
\begin{figure*}
	\centering
    \includegraphics[width=0.9\textwidth]{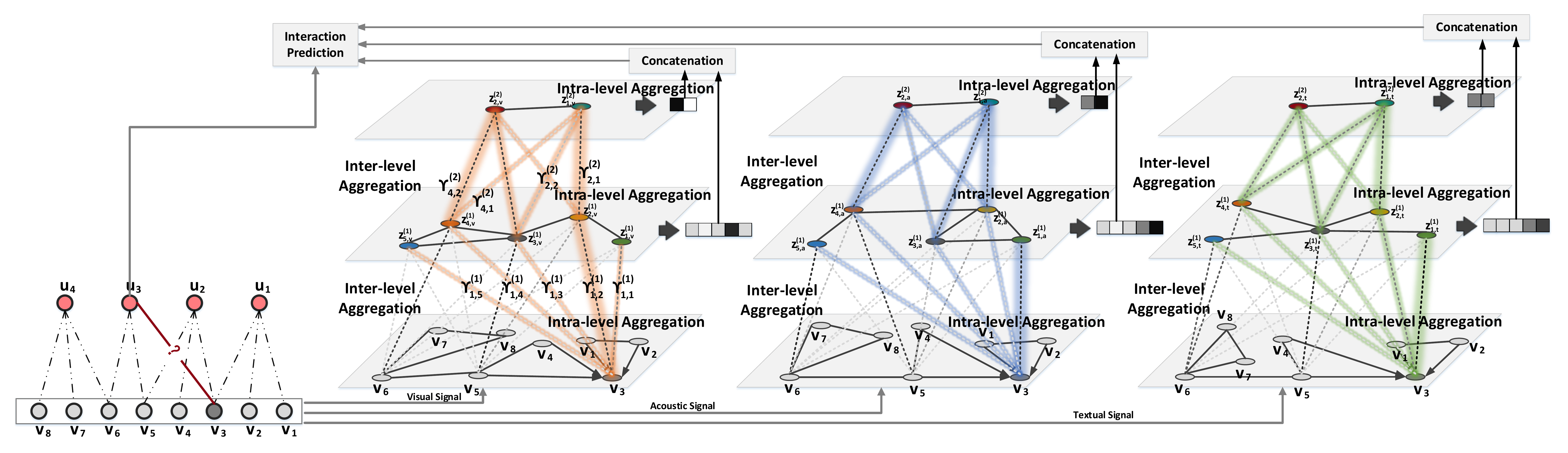}
    % \vspace{-15pt}
    \caption{Schematic illustration of our proposed model. It contains intra-level aggregation, cross-level aggregation, and interaction prediction operations. In the different modalities, the item is represented by its distributions over the learned user intents, which are remarked with orange, blue, and green dashed.}
	\label{fig:framework}
    \vspace{-10pt}
\end{figure*}
\section{Related Work}
In this section, we mainly review the work related to our research, including the multimedia recommendation and graph convolutional network.
\subsection{Multimedia Recommendation}
In the multimedia recommendation, items' content information is incorporated into the CF-based schema to generate a high-quality recommender system~\cite{CF1, CF2, CF3,CLCRec}. 
The fashions are based on the idea that the user preference on item contents can be modeled by their interaction historical records.

For this purpose, Chen~\textit{et al.}~\cite{ACF} explored the blurs of the item- and component-level feedback to the user preference profiling and proposed an attentive collaborative filtering~(ACF) module. 
Leveraging the proposed model, they learned the user preference from two levels of item features. 
However, due to neglecting users' effect during the feature extraction, the item features are unsuitable for user preference modeling.
Wei~\textit{et al.}~\cite{MMGCN} suggested that the users' tastes were aware of the different modalities' content information and presented a novel model, dubbed Multi-modal Graph Convolution Network. 
By using the graph convolutional operations, the method captures the model-specific user preference and distills the item representations simultaneously. 
Although the learned user representation can be treated as user preference, they prefer to reflect the users' specific characters on the item's fine-grained features instead of the user intents. 

Due to the limitation, some researchers work on refining the item representation with the user characters. 
By introducing some user demographics from social platforms, Cui~\textit{et al.}~\cite{REDAR} designed a regularized dual-factor regression model to learn common attribute representations for users and items. 
With the social attributes, they integrated user characters into the item representations. 
Further, Fan~\textit{et al.}~\cite{GraphRec} replaced the demographics with users' social network information and proposed a novel graph neural network framework for social recommendations, in which the social relation between users is combined with user-item interactions for features modeling. 
But, either the user profiles or the social information of users, is probably inaccessible, even irrelevant to the user preference.

The studies tend to learn the representations of user intents according to the supervisory signal. However, they ignore the complicate relation between the user intents, which facilitates the intents modeling. In this work, we consider the correlation between the fine-grained and coarse-grained intents as well as that between intents in the same level. As such, we organize the user intents associated with these two kinds of correlation as a hierarchical graph and utilize the graph convolutional operations to learn the representations of multi-level user intents.
\subsection{GCNs based Recommendation}
Empowered by the success of GCNs in the node representing, it is widely used by several tasks with non-Euclidean data, like computer vision~\cite{MM_cv_1,MM_cv_2}, natural language processing~\cite{MM_nlp_1,MM_nlp_2}, and recommender system~\cite{MM_rec_1,MM_rec_2,PHRec}.  
The core of the GCNs is collecting the signals of the node's neighbors and encoding the local structure information into the node representation, which enhances its representativeness. 

For the personalized recommendation, Wang~\textit{et al.}~\cite{NGCF} modeled the high-order connectivity in the user-item graph and explicitly injected the corresponding collaborative filtering signals into the user and item representation. 
Besides encoding the local structural information into the representation, GCNs operations can be used to learn hierarchical representations of graphs. 
For instance, Ying~\textit{et al.}~\cite{DiffPool} presented a differentiable graph pooling model to generate the hierarchical representations of graphs. 
Towards the goal, the module recursively leverages GCNs operations to learn a matrix, which is used to project the nodes into a coarser space. 
Similarly, Yu~\textit{et al.}~\cite{LGR} designed a general Layout-Graph Reasoning (LGR) layer to mine contextual graph semantic information from specific nodes to abstraction nodes. 
Although we follow the same idea to construct a hierarchical graph, we design two kinds of operations: (1) intra-level aggregation that distills the informative signal from item features to model the fine-grained intents; and (2) inter-level aggregation which explicitly represents the supernodes corresponding to coarse-grained intents in the higher level by summarizing the low-level nodes according to their affinities. 
\section{Methodology}
\subsection{Preliminary}
Given a set $\mathcal{U}$ of $N$ users, a set $\mathcal{V}$ of $M$ items, and their interaction matrix $\mathbf{R}\in \mathbb{R}^{N\times M}$, we use $\mathbf{R}_{i,j}$ to represent the interaction between user $\text{u}_i \in \mathcal{U}$ and item $\text{v}_j \in \mathcal{V}$. Thereinto, $\mathbf{R}_{i, j}=1$ denotes that $\text{u}_i$ has interacted with $\text{v}_j$; otherwise, $\mathbf{R}_{i, j}=0$.
Moreover, the multimedia content information of each item is provided, involving the visual, acoustic, and textual modalities. 
And, we use annotations $v$, $a$, and $t$ to indicate them, respectively. 
For notational convenience, without particular clarification, we discard these modality indicators to elaborate on each component in the single modality and do the same operations on other modalities.

As mentioned before, our research objective is to learn the multi-level user intents and accordingly optimize the representations of users and items.  
Following the assumption that users with similar historical behaviors prefer the similar items~\cite{BPR}, we propose to exploit the correlation between items co-interacted by the same users to distill their fine-grained intents towards the items. 
Formally, it could be formulated as,
\begin{equation}
    \mathbf{z}_i = f(\mathbf{x}_i, \mathbf{x}_{i,1}, \mathbf{x}_{i,2}, \dots, \mathbf{x}_{i,N_i}),
    \label{eq1}
\end{equation}
where $\mathbf{x}_i$ and $\mathbf{z}_i$ denote the representation vectors of item $\text{v}_i$ and the distilled information. 
In addition, $\mathbf{x}_{i,\cdot}$ and $N_i$ represent the item which is co-interacted with $i$ and the number of the co-interacted items of item $i$.
Further, to learn the representation of each coarser-grained user intent, we iteratively aggregate the finer-grained ones according to their affinities to the coarser-grained.

For this purpose, we construct a co-interacted item graph, $\mathcal{G}=\{\mathbf{X}, \mathbf{A}\}$, to organize the users' behavior and items' content information together. 
Whereinto, $\mathbf{x}_i\in \mathbf{X}$ is initialized with the extracted features in different modalities~(\textit{e.g.,} visual, acoustic, and textual features).
$\mathbf{A} \in \mathbb{R}^{M*M}$ reflects the connections between the nodes according to their co-interacted records.The presence of the connection between each node and its neighbor signifies that they are historically interacted by several user in common, like the videos co-watched by same audiences and products co-purchased by same customers. Specifically, we collect the item-item pairs  according to their historically interactions and treat them as the neighbors. The adjacent matrix $\mathbf{A}$ is formally defined as
\begin{equation}
    \mathbf{A}_{i, j} = 
    \begin{cases}
            1 &\  if\ \text{v}_j\ \in\ \mathcal{N}(\text{v}_i),\\[2mm]
            0 &\  if\ \text{v}_j\ \not\in\ \mathcal{N}(\text{v}_i),
    \end{cases}
\end{equation}
where $\mathcal{N}(\text{v}_i)$ represents the set of neighbors of $\text{v}_i$. 

\subsection{Model Framework}
In this section, we detail our proposed framework. 
As shown in Figure~\ref{fig:framework}, the framework contains three components, including the intra-level aggregation, inter-level aggregation, and interaction prediction. 
By iteratively performing the intra-level and inter-level aggregation operations, we capture information on the user intent from the co-interacted item graph and explicitly model their hierarchical structure from the fine-grained to the coarse-grained.
We then represent the items based on them and learn the users' characters to predict the interactions between users and items.

\subsubsection{Intra-level Aggregation}
Different from the prior works which assume the items are independent with each other, we explicitly leverage the correlation between items to distill the informative signal from features. 
Thus, we reorganize the items as a co-interacted graph and cast their correlations to the nodes' local structure in this graph. 

Taking the advantage of the message aggregation mechanism in GCNs~\cite{SIGIR2, SIGIR3}, the local structure information~(\textit{i.e.} user behaviors) of each node is encoded and used to refine its features~(\textit{i.e} content information), which distills the intents of users who have clicked or watched it. 
Therefore, each node could be represented with the vector conveying the user intents towards the item. Accordingly, we reformulate Eq.\ref{eq1} as the intra-level aggregation function at the $l$-th layer and provide the matrix form as, 
\begin{equation}
    \mathbf{Z}^{(l)} = f(\mathbf{X}^{(l)}, \mathbf{A}^{(l)}),
\end{equation}
where $f(\cdot)$ means the intra-level aggregation function. 
$\mathbf{X}^{(0)}$ is set as $\mathbf{X}$ at initial iteration and $\mathbf{A}^{(0)}$ denotes the adjacent matrix of co-interacted graph $\mathbf{A}$. 
With the aggregation operations, we learn the topological structure of each node's neighborhood as well as the distribution of node features in the neighborhood~\cite{GraphSAGE}. 
The distribution of the neighborhood features can provide a strong signal for distilling the information on the user intents~\cite{GIN}. 

More specifically, in our proposed framework, the aggregation function could be flexibly implemented using different methods, such as GraphSAGE~\cite{GraphSAGE} and GCN~\cite{GCN}. 
Here, we utilize a light variant of GCNs, which removes nonlinearities and collapses weight in the aggregation phase, formally
\begin{equation}
    \mathbf{Z}^{(l)}=\mathbf{D}^{-\frac{1}{2}}\mathbf{A}^{(l)}\mathbf{D}^{-\frac{1}{2}}\mathbf{X}^{(l)},
\end{equation}
where $\mathbf{D}$ is the degree matrix of adjacent matrix $\mathbf{A}^{(l)}$.
\subsubsection{Inter-level Aggregation}
Intuitively, the coarse-grained user intents could be summarized from the fine-grained ones. 
Therefore, we develop a inter-level aggregation operation to explicitly model the relationship between the low-level~(\textit{i.e.} fine-grained) and the high-level~(\textit{i.e.} coarse-grained) user intents. 

In particular, we establish several supernodes, which can be regarded as coarser-grained user intents, to construct $(l+1)$-th level graph. Here, we define and randomly initialize the supernodes as, 
\begin{equation}
    \mathbf{X}^{(l+1)} = [\ \mathbf{x}^{(l+1)}_{1}, \cdots, \mathbf{x}^{(l+1)}_{k}, \cdots, \mathbf{x}^{(l+1)}_{K^{(l+1)}}\ ],
\end{equation}
where $\mathbf{x}^{(l+1)}_{k}$ and $K^{(l+1)}$ denote the representation of the $k$-th supernode and the number of the supernodes in $(l+1)$-th layer, respectively. 

In order to discover the relationship of intent pairs across the levels, we perform the inner product for each node and supernode pair to measure its affinity.  
It is formulated as: 
\begin{equation}
    e_{i,k} = {{\mathbf{z}_i}^{(l)}}^\intercal \cdot \mathbf{x}^{(l+1)}_k,\\[2mm]
\end{equation}
where $\mathbf{z}^{(l)}_i$ is the vector representing the $i$-th node in lower-level graph and $e_{i, k}$ is the score of affinity between the $i$-th node and $k$-th supernode.

Following this, the weights of each node assigned to supernodes are calculated by normalizing the scores,
\begin{equation}
    \mathbf{\gamma}^{(l+1)}_{i,k} = \frac{\text{exp} \ e_{i,k}}{\sum_{k'=1}^{K^{(l+1)}}{\text{exp}\ e_{i,k'}}},  
\end{equation}
where $\mathbf{\gamma}^{(l+1)}_{i,k}$ is the assignment weight of the $i$-th node towards $k$-th established supernode.  
Through such an assignment operation, our model constitutes the relationship across the lower level and higher level graphs. 

Beyond the relationship between the finer-grained and coarser-grained intent pair, the adjacent matrix of the high-level graph can be computed with the low-level nodes' adjacent matrix and assignment matrix. It is formulated as,
\begin{equation}
    \mathbf{A}^{(l)} = 
        \begin{cases}
            \mathbf{A}\  &\   l=0\ ,\\[3mm]
            {\mathbf{\Gamma}^{(l)}}^\intercal \ \mathbf{A}^{(l-1)}\ \mathbf{\Gamma}^{(l)}\  &\  otherwise,
        \end{cases}
\end{equation}
where $\mathbf{\Gamma}^{(l)}\in\mathbb{R}^{K^{(l-1)}*K^{(l)}}$ denotes the matrix of assignment weights. 
Whereinto, its $i$-th row $\bm{\gamma}_{i}$ is the assignment weight vector of $i$-th node over the supernodes. 
Considering that using graphs has the limitation of providing exclusive sets in the space partitioning~\cite{SB2}, we adopt a soft-assign strategy to alleviate the unnecessary overlapping. 
Thus, we introduce a cross-entropy loss to make each node's assignment vector close to the one-hot vector, which facilitates to disentangle the user intents in the same level.   
It is defined as,
\begin{equation}
    \label{Equ_CrossEntropy}
    \mathcal{L}^{(l)}_1 =  -\frac{1}{K^{(l)}}\sum_{k'=1}^{K^{(l)}}{\gamma^{(l)}_{i,k'}}^\intercal \ln{\gamma^{(l)}_{i,k'}}\ .
\end{equation}

Moreover, we introduce an independence loss to keep user intents independent, so as to enhance the robustness of the proposed model, formally,   
\begin{equation}
    \label{Equ_Independent}
    \mathcal{L}^{(l)}_2 = \frac{1}{K^{(l)}} \left\|\ {\mathbf{X}^{(l)}}^\intercal \mathbf{X}^{(l)} - \mathbf{I}^{(l)}\ \right\|_{2},
\end{equation}
where $\mathbf{I}^{(l)}\in\mathbb{R}^{K^{(l)}*K^{(l)}}$ is the identity matrix in the $l$-th level.
\subsubsection{Interaction Prediction}
By iteratively conducting $L$ intra-level and inter-level aggregation operations, our model explicitly models the user intents as well as their hierarchical structure. 
We then calculate the distributions over the user intents at each level for each item, as 
\begin{equation}
    \begin{cases}
        \hat{x}^{(1)}_i =  x^\intercal_i \Gamma^{(1)}, \\[2mm]
        \hat{x}^{(2)}_i = \hat{x}^{(1)}_i \Gamma^{(2)}, \\[2mm]
        % \hat{x}^{(3)}_i = \hat{x}^{(2)}_i \Gamma^{(3)} , \\[2mm]
        \cdots, \\[2mm]
        \hat{x}^{(L)}_i = \hat{x}^{(L-1)}_i \Gamma^{(L)},     
    \end{cases}
\end{equation}
where $\hat{x}_i^{(l)}$ represents the item according to the user intents at the $l$-th layer.  

Considering the vectors are based on the user intents in different levels, we employ the layer-aggregation mechanism~\cite{KGAT,NGCF} to integrate these vectors as the item's representation, formally,  
\begin{equation}
    \mathbf{v}_i = \hat{x}^{(1)}_i\ ||\ \hat{x}^{(2)}_i\ ||\ \cdots\ ||\ \hat{x}^{(L)}_i\ ,
\end{equation}
where the symbol $||$ means the concatenation operation.

Then, we define and randomly initialize the trainable representations of users, formally, 
\begin{equation}
    \mathbf{u}_j = [\ \underbrace{\mathbf{u}_{j,1}, \dots, \mathbf{u}_{j, K^{(1)}}}_\text{1-st level},\cdots , \underbrace{{\mathbf{u}_{j, 1}},\cdots,\mathbf{u}_{j,K^{(L)}}}_\text{L-th level}\ ],
\end{equation}
where $\mathbf{u}_{i, \cdot}$ could be used to measure the user's characteristics on the corresponding intent. 
These representation vectors are learned by the back-propagation operation during the training phase. 

Following the idea that users have different preferences in different modalities~\cite{MMGCN}, we combine the multi-modal representations in different modalities as a whole vector, formally,   
\begin{equation}
    \begin{cases}
        \mathbf{u}^{*}_i = \ \mathbf{u}^{v}_i\ +\ \mathbf{u}^{a}_i\ +\ \mathbf{u}^{t}_i \ ,\\[2mm]
        \mathbf{v}^{*}_j = \ \mathbf{v}^{v}_j\ +\ \mathbf{v}^{a}_j\ +\ \mathbf{v}^{t}_j \ ,
    \end{cases}
\end{equation}
where $\mathbf{u}^{*}_i$ and $\mathbf{v}^{*}_i$ denote the user's and item's multi-modal representations, respectively. 
\subsection{Optimization}
To learn the parameter of the proposed model, we concatenate the ID embedding vectors and representation vectors for the users and items, as 
\begin{equation}
        \Bar{\mathbf{u}}_i = \ \hat{\mathbf{u}}_{i}\ ||\ \mathbf{u}^{*}_i \ ,\ \ \  
        \Bar{\mathbf{v}}_j = \ \hat{\mathbf{v}}_{j}\ ||\ \mathbf{v}^{*}_j \ ,
\end{equation}
where $\hat{\mathbf{u}}_{i}$ and $\hat{\mathbf{v}}_{j}$ denote the ID embedding vectors of the user and item, respectively.  
Then, Bayesian Personalized Ranking~(BPR)~\cite{BPR} is adopted to perform the pair-wise ranking for the recommendation. 
As such, we construct a triplet of one user $\text{u}_i$, one observed item $\text{v}_p$, and one unobserved item $\text{v}_q$, formally as, 
\begin{equation}
    \mathcal{T} = \{(\text{u}_i, \text{v}_p, \text{v}_q)\ | \mathbf{R}_{i,p}=1,\ \mathbf{R}_{i, q}=0 \},
\end{equation}
where $\mathcal{T}$ is a triplet set for training. 
The BPR loss function can be defined as, 
\begin{equation}
    \mathcal{L}_3 = \sum_{(\text{u}_i, \text{v}_p, \text{v}_q)\in\mathcal{T}}{-\ln \phi(\Bar{\mathbf{u}}_i^\intercal\Bar{\mathbf{v}}_p - \Bar{\mathbf{u}}_i^\intercal\Bar{\mathbf{v}}_q)},
\end{equation}
where $\phi(\cdot)$ is the sigmoid function. 

Combining with the cross-entropy loss and independence loss, we reach the objective function as follows, 
\begin{equation}
    \mathcal{L}\ =\ \lambda_1\sum_{l=1}^{L}\mathcal{L}_1\ +\ \lambda_2\sum_{l=1}^{L}\mathcal{L}_2\  +\ \mathcal{L}_3\ +\ \lambda_3||\theta||_2 ,
\end{equation}
where $\lambda_1, \lambda_2, \lambda_3$ and $\theta$ represent the regularization weights and the parameters of the model, respectively. 
\subsection{Discussions}
\label{discuss}
\subsubsection{Model Size.}
Although $L$ layers are stacked to model the multi-level graph structure, there are few additional parameters introduced into our framework. 
Specifically, in each aggregation layer, we simplify the aggregating operations, in which the nonlinear function and weight matrices are discarded.  
Therefore, it does not introduce any external parameter for computing. 
In the inter-level aggregation layer, to represent the user intents in various layers, we establish several supernodes and represent them with the vectors. 
In particular, we initialize $K^{(l)}$ vectors of $D$ dimensions in the $l$-th layer, where $D$ and $K^{(l)}$ are much less than the number of items.  
In addition, considering that $L$ is usually a number smaller than 5, the cost of these external parameters is negligible. 
To summarize, our algorithm uses very few additional model parameters to learn the multi-level user intents and predict the interactions between users and items. 
\subsubsection{Complexity Analysis.} 
According to the architecture of our proposed framework, we elaborate on each component to analyze the time cost.
In particular, benefiting from the light variant of GCN, the $l$-th intra-level aggregation layer has computational complexity $\mathcal{O}(K^{(l)}\times K^{(l)}\times D)$ without any other cost, such as the cost from the element-wise nonlinear operation and matrix multiplication. 
In terms of the $l$-th inter-level aggregation layer, its computational complexity is $\mathcal{O}(K^{(l)}\times K^{(l-1)}\times D)$. 
Even though there are multiple layers in our framework, the computational complexity closes to $\mathcal{O}(M\times M\times D)$, since $\sum^{L}_{l=1}{K^{(l)}}$ is much smaller than $M$. 
In addition, only the inner product is conducted in the prediction layer, for which the time cost of the whole training epoch is $\mathcal{O}(|\mathcal{T}|\times D)$. Whereinto, $|\mathcal{T}|$ is the number of triplets in the training set.  
Therefore, the overall training complexity of our proposed model is closed to $O(M\times M\times D+|\mathcal{T}|\times D)$, while the complexity of MMGCN is $O(|\mathcal{T}|\times D+(M+N)\times (M+N)\times D)$. 

\section{EXPERIMENTS}
\begin{table}
\small
  \centering
  \caption{Summary of the datasets. The dimensions of visual, acoustic, and textual modalities are denoted by V, A, and T, respectively.}
    %   \vspace{-5pt}
  \label{table_1}
  \setlength{\tabcolsep}{1.0mm}
  \begin{tabular}{c|c|c|c|c|c|c}
    \specialrule{0.1mm}{0pt}{1pt}
    Dataset& \#Interactions & \#Items & \#Users & V & A & T\\
    \specialrule{0.1mm}{1pt}{1pt}
    \specialrule{0.1mm}{1pt}{1pt}
    Movielens&1,239,508&5,986&55,485&2,048&128&100\\
    \specialrule{0.0mm}{0pt}{1pt}
    Tiktok&726,065&76,085&36,656&128&128&128\\
    \specialrule{0.0mm}{0pt}{1pt}
    Kwai&298,492&86,483&7,010&2,048&-&-\\
    % \hline
    \hline
  \end{tabular}
\end{table}
\begin{table*}
  \centering
  \renewcommand\arraystretch{1.2}
  \setlength{\tabcolsep}{2.8mm}
  \caption{Performance comparison between our model and the baselines over the three datasets.}
%   \vspace{-5pt}
  \label{table_2}
  \begin{tabular}{c|c|c|c|c|c|c|c|c|c|c}

    \specialrule{0.1mm}{1pt}{1.5pt}  
    \multicolumn{2}{c|}{Model}&VBPR&DUIF&CB2CF&NGCF&MMGCN&DisenGCN&MacridVAE&Ours&\%Improv.\\
    \hline
    \hline
    \multirow{9}{*}{Movilens}
    &P@1&0.0819&0.0864&0.0915&0.0863&\underline{0.0982}&0.0904&0.0935&\textbf{0.1130}&\textbf{15.07\%}\\
    &P@5&0.0587&0.0639&0.0664&0.0635&\underline{0.0696}&0.0660&0.0640&\textbf{0.0790}&\textbf{13.51\%}\\
    &P@10&0.0512&0.0538&0.0548&0.0547&\underline{0.0581}&0.0555&{0.0576}&\textbf{0.0619}&\textbf{7.47\%}\\
    &R@1&0.0368&0.0382&0.0406&0.0354&\underline{0.0436}&0.0400&0.0421&\textbf{0.0496}&\textbf{13.77\%}\\
    &R@5&0.1258&0.1334&0.1417&0.1071&\underline{0.1513}&0.1415&0.1501&\textbf{0.1649}&\textbf{8.99\%}\\
    &R@10&0.1990&0.2167&0.2265&0.2196&\underline{0.2345}&0.2222&0.2286&\textbf{0.2522}&\textbf{10.32\%}\\
    &NGCG@1&0.0819&0.0864&0.0915&0.0863&\underline{0.0982}&0.0904&0.0935&\textbf{0.1130}&\textbf{15.07\%}\\
    &NGCG@5&0.1648&0.1818&0.1881&0.1840&\underline{0.2000}&0.1911&0.1898&\textbf{0.2258}&\textbf{12.90\%}\\
    &NDCG@10&0.2261&0.2341&0.2505&0.2342&\underline{0.2517}&0.2401&0.2437&\textbf{0.2677}&\textbf{9.35\%}\\
    \hline
    \multirow{9}{*}{Tiktok}
    &P@1&0.0161&0.0043&0.0156&0.0174&0.0168&0.0172&\underline{0.0181}&\textbf{0.0189}&\textbf{4.42\%}\\
    &P@5&0.0106&0.0050&0.0116&0.0139&0.0150&0.0154&\underline{0.0160}&\textbf{0.0171}&\textbf{6.88\%}\\
    &P@10&0.0118&0.0087&0.0109&0.0135&0.0144&0.0145&\underline{0.0152}&\textbf{0.0164}&\textbf{7.89\%}\\
    &R@1&0.0098&0.0015&0.0104&0.0117&0.0102&0.0120&\underline{0.0132}&\textbf{0.0138}&\textbf{4.55\%}\\
    &R@5&0.0364&0.0115&0.0374&0.0383&\underline{0.0428}&0.0420&\underline{0.0428}&\textbf{0.0479}&\textbf{11.92\%}\\
    &R@10&0.0628&0.0483&0.0642&0.0780&0.0808&0.0760&\underline{0.0813}&\textbf{0.0884}&\textbf{11.48\%}\\
    &NGCG@1&0.0161&0.0043&0.0156&0.0174&0.0168&0.0172&\underline{0.0181}&\textbf{0.0189}&\textbf{4.42\%}\\
    &NGCG@5&0.0306&0.0282&0.0362&0.0405&0.0445&0.0412&\underline{0.0472}&\textbf{0.0520}&\textbf{10.17\%}\\
    &NDCG@10&0.0574&0.0434&0.0613&0.0661&0.0674&0.0639&\underline{0.0686}&\textbf{0.0769}&\textbf{12.10\%}\\
    \hline
    \multirow{9}{*}{Kwai}
    &P@1&0.0342&0.0204&0.0367&\underline{0.0385}&0.0381&0.0353&0.0382&\textbf{0.0387}&\textbf{0.52\%}\\
    &P@5&0.0174&0.0124&\underline{0.0177}&0.0159&0.0142&0.0160&0.0162&\textbf{0.0185}&\textbf{4.52\%}\\
    &P@10&\underline{0.0132}&0.0095&\underline{0.0132}&0.0118&0.0120&0.0127&{0.0130}&\textbf{0.0140}&\textbf{6.07\%}\\
    &R@1&0.0116&0.0062&0.0152&0.0167&\underline{0.0176}&0.0164&0.0175&\textbf{0.0183}&\textbf{3.98\%}\\
    &R@5&\underline{0.0284}&0.0175&0.0282&0.0278&0.0284&0.0270&0.0282&\textbf{0.0307}&\textbf{8.10\%}\\
    &R@10&\underline{0.0410}&0.0258&0.0407&0.0402&0.0398&0.0403&{0.0405}&\textbf{0.0434}&\textbf{5.85\%}\\
    &NGCG@1&0.0342&0.0204&0.0367&\underline{0.0385}&0.0381&0.0353&0.0382&\textbf{0.0387}&\textbf{0.52\%}\\
    &NGCG@5&\underline{0.0576}&0.0493&\underline{0.0576}&0.0532&0.0556&0.0512&0.0512&\textbf{0.0641}&\textbf{11.30\%}\\
    &NDCG@10&\underline{0.0702}&0.0521&0.0700&0.0699&0.0681&0.0683&{0.0685}&\textbf{0.0778}&\textbf{10.70\%}\\
    \specialrule{0.1mm}{1pt}{1.5pt}  
  \end{tabular}
\end{table*}

To evaluate the proposed model, we conducted experiments on the three public datasets to answer the following research questions:
\begin{itemize}
    \item \textbf{RQ1} How does our proposed model perform compared with state-of-the-art multimedia recommendation models and other user intent modeling based methods?
    \item \textbf{RQ2} How does each design (\textit{i.e.} structure of user intents, number of the modalities, cross-entropy loss and independence loss)  affect the performance of our model?
    \item \textbf{RQ3} Can our model provide the semantic descriptions for user intents? 
\end{itemize}

Before answering the above three questions, we first describe the datasets, evaluation metrics, baselines, and parameter settings in the experiments.
\subsection{Experiments Settings}

\subsubsection{Dataset}
To evaluate our model, extensive experiments are conducted on three publicly accessible datasets, namely Movielens, Tiktok, and Kwai, which are designed for the micro-video personalized recommendation. 
The statistics of datasets are summarized in Table~\ref{table_1}. 
\begin{itemize}
    \item \textbf{Movielens}
    To support the multimedia recommendation, Movielens dataset\footnote{https://movielens.org/.} has been extended by collecting the videos' trailers, titles, and descriptions~\cite{MMGCN}. 
    In order to learn the visual and acoustic features, the keyframes and audio track are extracted from the trailers. 
    With the pre-trained models~(\textit{e.g.} ResNet~\cite{ResNet}, VGGish~\cite{VGG}, and Sentance2Vector~\cite{S2V}) , the visual, acoustic, and textual features are extracted from the frames, audio tracks, and descriptions, respectively. 
    \item \textbf{Tiktok} 
    This dataset is obtained from Tiktok\footnote{https://www.tiktok.com/.}, which is a popular micro-video sharing platform. 
    It consists of users, micro-videos, and their historical interactions. 
    Besides, the multi-modal features are also extracted from the micro-videos, involving visual, acoustic, and textual modalities.
    \item \textbf{Kwai} 
    As a micro-video service provider, Kwai\footnote{https://www.kwai.com/.} released a large-scale micro-video dataset for user behavior prediction. 
    Besides the users and micro-videos information, the dataset contains the users behaviors towards the micro-video~(\textit{i.e.} clicks, likes, and follows) associated with the timestamps. 
    Within this dataset, we collected some interaction records during a certain period for our experiments. 
    Different from the above datasets, the audio and textural information is missing in Kwai. 
\end{itemize}

For each dataset, we randomly split the historical interaction records into the training set, validation set, and testing set with the ratio $8:1:1$. 
Whereinto, the validation set and testing set are respectively used to tune the hyper-parameters and evaluate the performance in the experiments.  
With regard to the training set, we performed negative sampling to create the training triples and constructed the co-interacted item graph for our proposed model. 
Specifically, we captured the video pairs which were co-watched by at least 5 different users. 
We then treated obtained videos as the nodes of the graph and connected them corresponding to the co-watched video pairs. 
\subsubsection{Evaluation Metrics}
For each user in the validation set and the testing one, we regarded the micro-videos which the user had no interaction with as the negative samples. 
Using the trained model, we scored the interactions between the user and micro-video pairs and ranked them in the descending order. 
In addition to the precision@K~(P@K for short) and recall@K~(R@K for short), we also adopted Normalized Discounted Cumulative Gain~(NDCG@K for short) to evaluate the effectiveness of top-K recommendation. 
% Specifically, it is formally defined as:
% \begin{equation}
%     NDCG@K=\frac{\text{DCG}_K}{\text{IDCG}_K},\ 
%     \text{and DCG}_K=\sum\limited_{i=1}^K{\frac{2^{r_{ui}}-1}{\log_2(1+i)}}, \nonumber
% \end{equation}
% where IDCG is ideal DCG, and $r_{ui}$ denotes the interaction between user $u$ and item $i$. 
By default, we set $K=10$ and reported the average values of the above metrics for all users during the testing phase. 

\subsubsection{Baselines}
To evaluate our proposed model, we compared it with several state-of-the-art baselines. 
We briefly divided these baselines into three groups: 1) deep-learning-based methods~(\textit{i.e., VBPR, DUIF, CB2CF}), 2) user-intent-agnostic methods~(\textit{i.e., NGCF, MMGCN}), and 3) user-intent-based methods~(\textit{i.e., DisenGCN, MacridVAE}).
\begin{itemize}
    \item \textbf{VBPR}~\cite{VBPR}
    This is a benchmark model in the multimedia recommendation. 
    It incorporates the content information into the collaborative filtering framework. 
    To ensure fairness, we fused the multi-modal features as the side information and fed them into the model, which could infer user preference towards the learned features. 
    \item \textbf{DUIF}~\cite{DUIF}
    Different from the CF-based model, this method learns the user reference to the item's features, which are concatenated and refined by the trainable deep learning model.
    \item \textbf{CB2CF}~\cite{CB2CF}
    It performs a deep neural multiview model to represent the rich content information. And, a mean squared error~(MSE) serves as a bridge from items' feature representations into their collaborative embeddings. 
    \item \textbf{NGCF}~\cite{NGCF}
    Upon the user-item bipartite graph, the method propagates the users and items representations to model the high-order connectivity in the graph. 
    It leverages the collaborative filtering signals to explicitly encode user information into item representations. 
    \item \textbf{MMGCN}~\cite{MMGCN}
    The model is a novel graph based framework designed for the multimedia recommendation, in which the model-specific user preference can be learned. 
    Performing the graph convolutional operations on the user-item graph in each modality, it can refine the items' representations with the learned users' model-specific preferences. 
    \item \textbf{DisenGCN}~\cite{DisenGCN}
    It is proposed for node predictions, in which the nodes' representations are disentangled into independent factors. 
    To model the user intents hidden in the representations of users and items, we extended it into the personalized recommendation task. 
    As such, we constructed the user-item graph and disentangled the nodes' representations into several independent channels, which can be viewed as the representations of user intents. 
    Based on this model, the user intents are captured and incorporated into the item representations. 
    \item \textbf{MacridVAE}~\cite{DisenVAE}
    This model aims to disentangle the user intents into the macro- and micro- level. 
    Thus, some concepts are learned to represent the macro-level intents, each dimension of which is viewed as the users' micro-level intent. 
    To model the items with learned user intents, it forces each item associated with a unique concept and represents it with the corresponding one-hot vector. 
\end{itemize}
\begin{table*}
 \renewcommand\arraystretch{1.2}
  \centering
  \setlength{\tabcolsep}{3.3mm}
  \caption{Performance of Proposed Model w.r.t. the structure of user intents over three datasets. (Each constant in the first column denotes the number of user intents in each layer, and the number of constants is the number of layers.)}
  \label{table_3}
  \begin{tabular}{c|ccc|ccc|ccc}
    \hline
    \multirow{2}{*}{Layer}&\multicolumn{3}{c|}{Movielens}&\multicolumn{3}{c|}{Tiktok}&\multicolumn{3}{c}{Kwai}\\
    &Precision&Recall&NDCG&Precision&Recall&NDCG&Precision&Recall&NDCG\\
    \hline
    \{32\}&0.0617&0.2502&0.2660&0.0161&0.0861&0.0766&0.0130&0.0403&0.0724\\
    \{32, 8\}&0.0617&0.2509&0.2660&0.0157&0.0845&0.0748&0.0138&0.0414&0.0760\\
    \{32, 8, 4\}&\textbf{0.0619}&\textbf{0.2522}&\textbf{0.2677}&\textbf{0.0164}&\textbf{0.0884}&\textbf{0.0769}&\textbf{0.0140}&\textbf{0.0434}&0.0778\\
    \{64, 16, 4\}&0.0618&0.2517&0.2675&0.0153&0.0830&0.0733&\textbf{0.0140}&0.0428&\textbf{0.0787}\\
    \{64, 32, 8, 4\}&0.0616&0.2512&0.2666&0.0160&0.0865&0.0763&0.0135&0.0421&0.0747\\
    \hline
  \end{tabular}
\end{table*}

\subsubsection{Parameter Settings}
We implemented our model with the help of Pytorch\footnote{https://pytorch.org/.} and torch-geometric package\footnote{https://pytorch-geometric.readthedocs.io/.}. 
More specifically, we initialized the model parameter with the Xavier~\cite{Xavier} initializer and took the Adam~\cite{Adam} as the optimizer. 
And, to ensure the fair comparison, we fixed the dimension of the embedding vector to 64 for all models.  
In terms of the hyperparameters, we used the grid search: 
the learning rate is tuned in $\{0.0001, 0.001, 0.01, 0.1, 1\}$ and regularization weight is searched in $\{0.00001, 0.0001, 0.001, 0.01, 0.1\}$. 
Besides, we employed the early stopping strategy, which stops the training if recall@10 on the validation data does not increase for 20 successive epochs. 
For the baselines, we did the same options and followed the designs in their articles to achieve the best performance. 

\subsection{Performance Comparison~(RQ1)}
To evaluate our proposed model, we report the empirical results of all methods and the improvements, which are calculated between our proposed method and the strongest baselines highlighted with underline, in Table~\ref{table_2}.
Analyzing this table from top to bottom, we observed:
\begin{itemize}
    \item 
    On Movielens and Tiktok, deep-learning-based models, especially DUIF purely based on the features, have poor performance~\textit{w.r.t.} precision, recall, and NDCG.
    It demonstrates that the features extracted by the pre-trained model are suboptimal in the multimedia recommendation. 
    Although VBPR and CB2CF achieve the comparable results on Kwai, we suggest that it may benefit from the dense interactions per user. 
    \item
    In most cases, the methods encoding the user information into item representations outperform the user-agnostic based model. 
    From this comparison, it indicates that the user characteristics enhance the item representations. 
    With users' preferences, the item representations are probably refined to reflect the features users concerned. 
    \item
    Compared with NGCF and MMGCN, the user intent modeling based approaches, including DisenGCN, MarcridVAE, and our proposed model, achieve better results. 
    It demonstrates that representing users and items with user intents boosts the personal recommendation. 
    This is because the user intents reflect the users' motivations and tastes. 
    Whereas, user preference learned by NGCF and MMGCN is used to filter out the non-essential features in the recommendation. 
    \item 
    As we can see from the results, there is a gap between the performance of DisenGCN and MarcridVAE. 
    We believe that the gap is caused by the different levels of user intents. 
    Compared with the single-level user intents, the two-level intents learned by MarcridVAE could be treated as an extension, which further disentangles the user intents into the specific ones and general ones. 
    \item 
    Without any doubt, our proposed model significantly outperforms the state-of-the-art baselines. 
    In terms of the NDCG@10, it improves over the strongest baselines by $9.35\%$, $12.10\%$, and $10.7\%$. 
    We suggest that the improvement mainly comes from the multi-level user intents as well as the user and item representations based on them.  
    Different from DisenGCN, our model constructs the hierarchical structure to model the multi-level user intents and explicitly learns the user intents representation.  
    Besides, our model uses the continuous vector to characterize both users and items, which alleviates the restriction of MacridVAE on the item representation.  
    Moreover, observing the results of top-K recommendation lists where K is in range of \{1, 5, 10\}, they show that our proposed model is robust to the various recommendation scenes and achieves the state-of-the-art performance on three datasets. 
    
\end{itemize}

\subsection{Ablation Study~(RQ2)}
\subsubsection{Effects of User intents' Structure}
In this work, we focus on modeling user intents as well as their hierarchical structure to optimize the user and item representations. 
To investigate the effect of structure, we conducted experiments on different user intents structures. 
In particular, we searched the number of layers in the range of $\{1, 2, 3, 4\}$. 
Also, we considered the comparison between the structures, which contain the same layers but different numbers of user intents in each layer. 
\begin{table}
  \centering
  \renewcommand\arraystretch{1.2}
  \setlength{\tabcolsep}{0.7mm}
  \caption{Performance of HUIGN with different modalities on Movielens and Tiktok datasets. The visual, acoustic, textual, and multiple denote the visual, acoustic, textual, and multiple modality representations, respectively.}
  \label{table_4}
  \begin{tabular}{c|ccc|ccc}
    \hline
    \multirow{2}{*}{Modality}&\multicolumn{3}{c|}{Movielens}&\multicolumn{3}{c}{Tiktok}\\
    &Precision&Recall&NDCG&Precision&Recall&NDCG\\
    \hline
    \specialrule{0.1mm}{1pt}{1.5pt}
    Visual&\textbf{0.0631}&\textbf{0.2555}&\textbf{0.2710}&0.0145&0.0796&0.0706\\
    \specialrule{0.0mm}{0pt}{1pt}
    Acoustic&0.0616&0.2510&0.2671&0.0144&0.0788&0.0694\\
    \specialrule{0.0mm}{0pt}{1pt}
    Textual&0.0611&0.2491&0.2621&0.0142&0.0770&0.0675\\
    \specialrule{0.1mm}{1pt}{1.5pt}        
    \specialrule{0.1mm}{0pt}{1pt}
    Multiple&{0.0619}&{0.2522}&{0.2677}&\textbf{0.0164}&\textbf{0.0884}&\textbf{0.0769}\\
    \hline
  \end{tabular}
\end{table}

\begin{figure*}
    \centering
    \subfigure[Precision@10 on Movielens]{
      \includegraphics[width=0.3\textwidth]{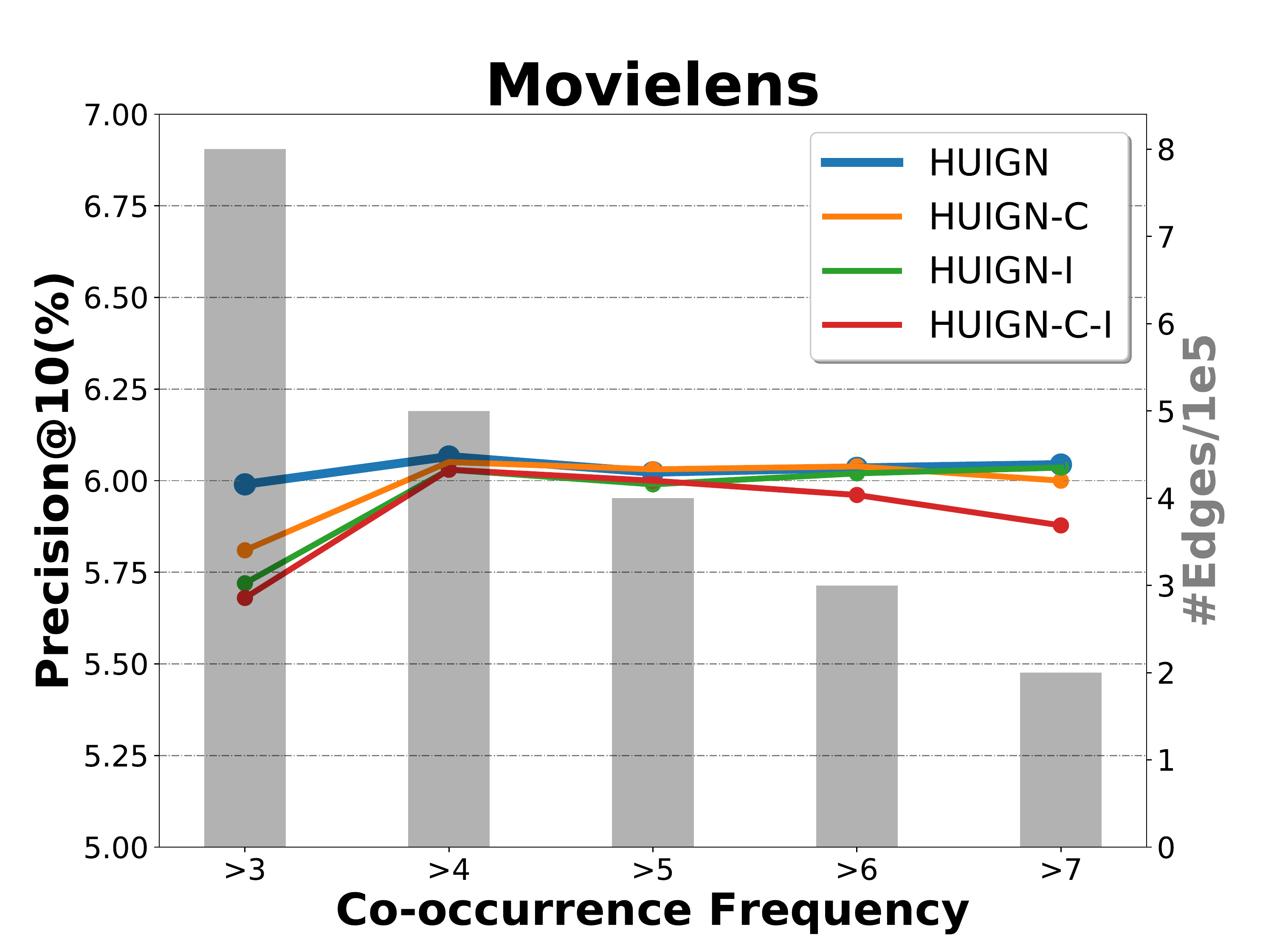}
      \label{fig_visualize_1_1}
    }
    \subfigure[Recall@10 on Movielens]{
      \includegraphics[width=0.3\textwidth]{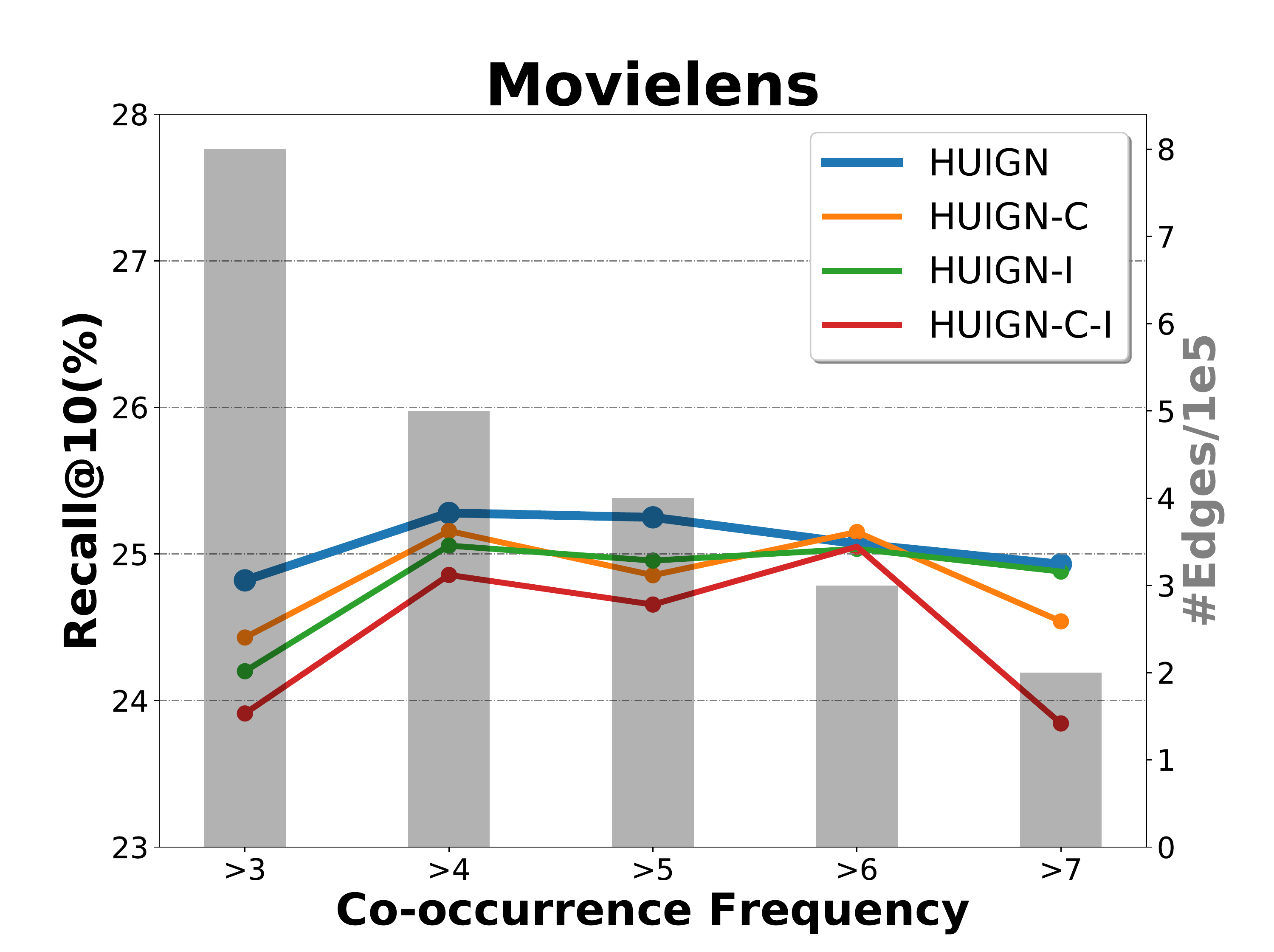}
      \label{fig_visualize_1_2}
    }
    \subfigure[NDCG@10 on Movielens]{
      \includegraphics[width=0.3\textwidth]{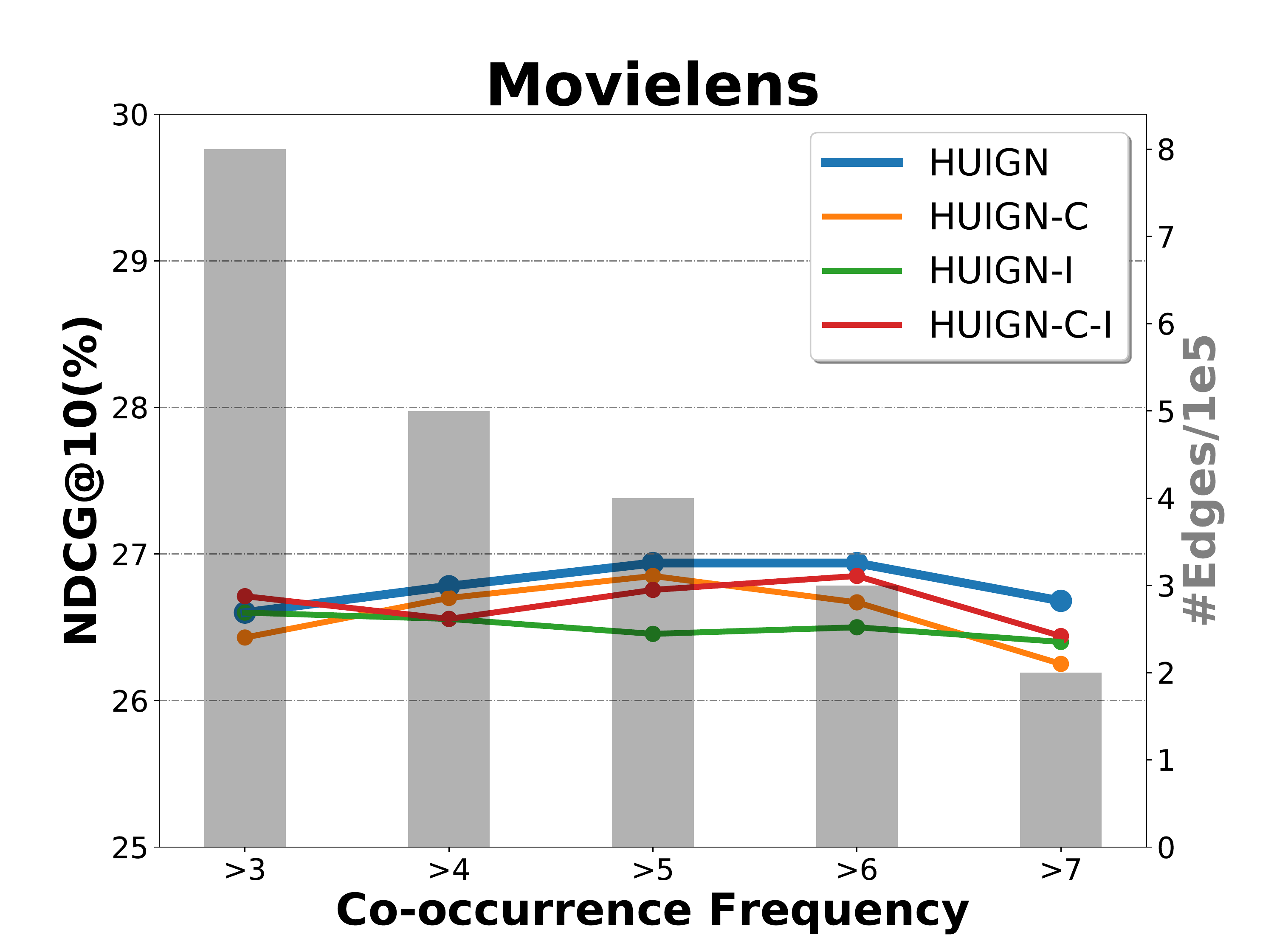}
      \label{fig_visualize_1_3}
    }
    \subfigure[Precision@10 on Kwai]{
      \includegraphics[width=0.3\textwidth]{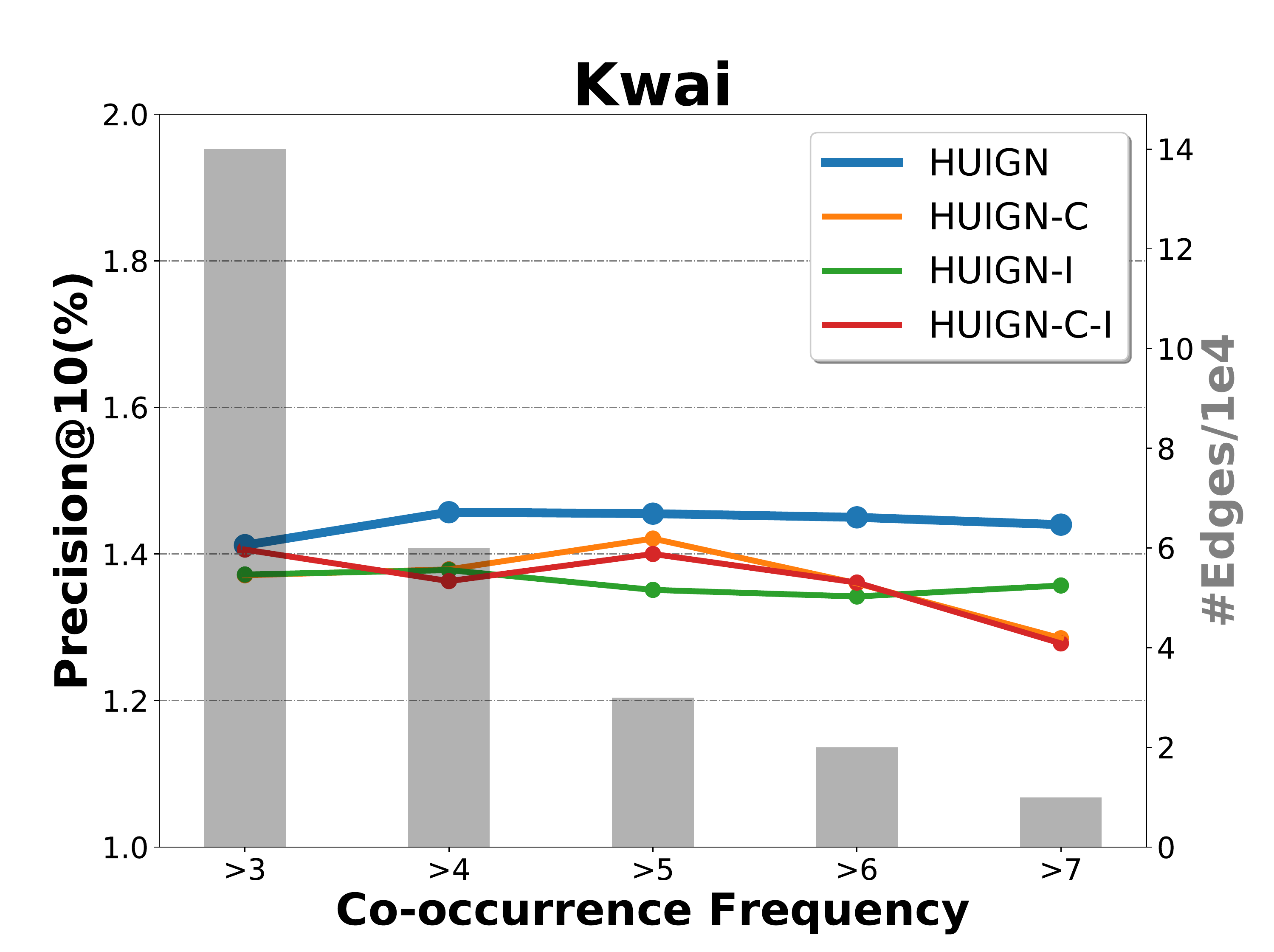}
      \label{fig_visualize_2_1}
    }
    \subfigure[Recall@10 on Kwai]{
      \includegraphics[width=0.3\textwidth]{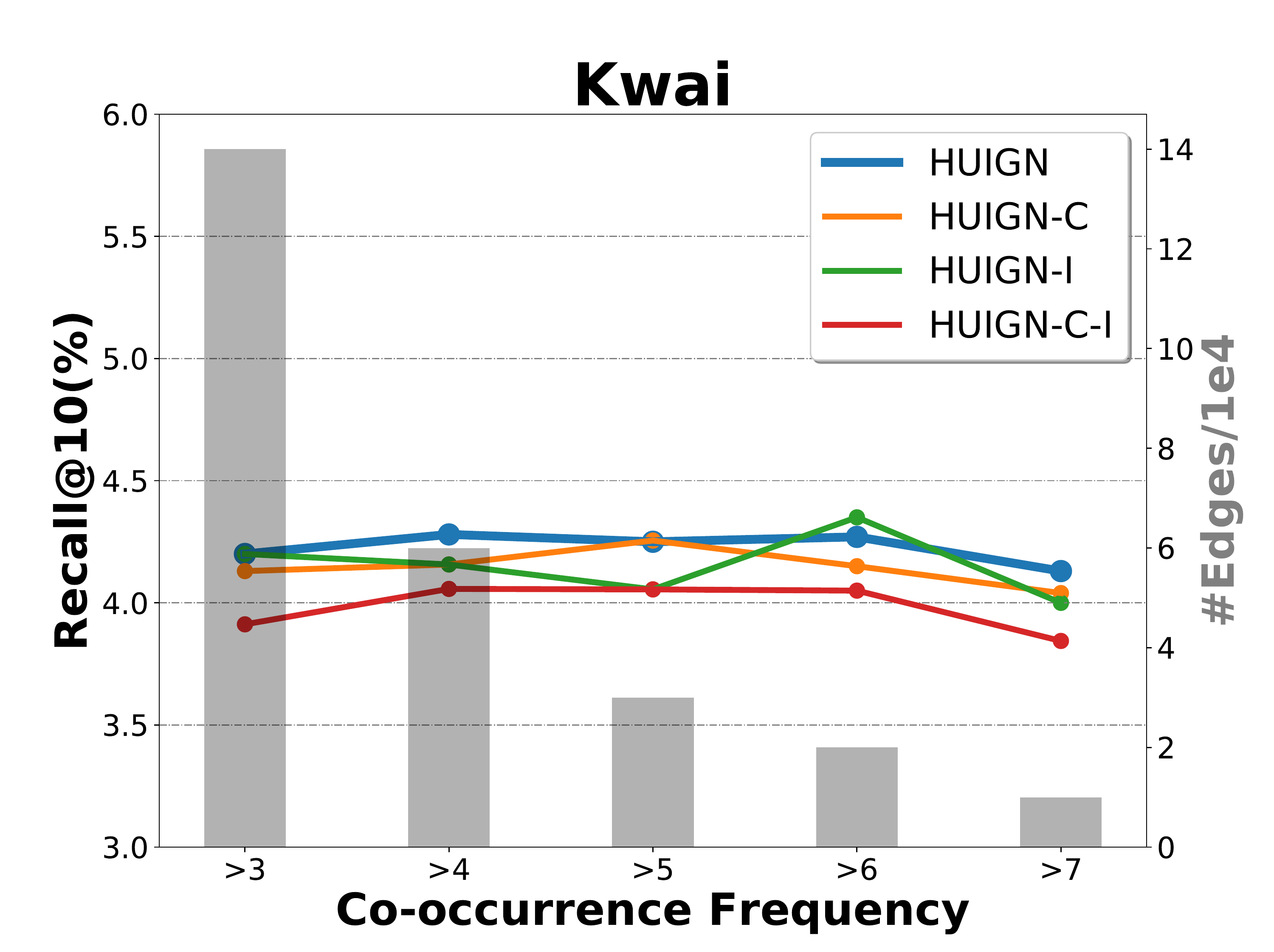}
      \label{fig_visualize_2_2}
    }
    \subfigure[NDCG@10 on Kwai]{
      \includegraphics[width=0.3\textwidth]{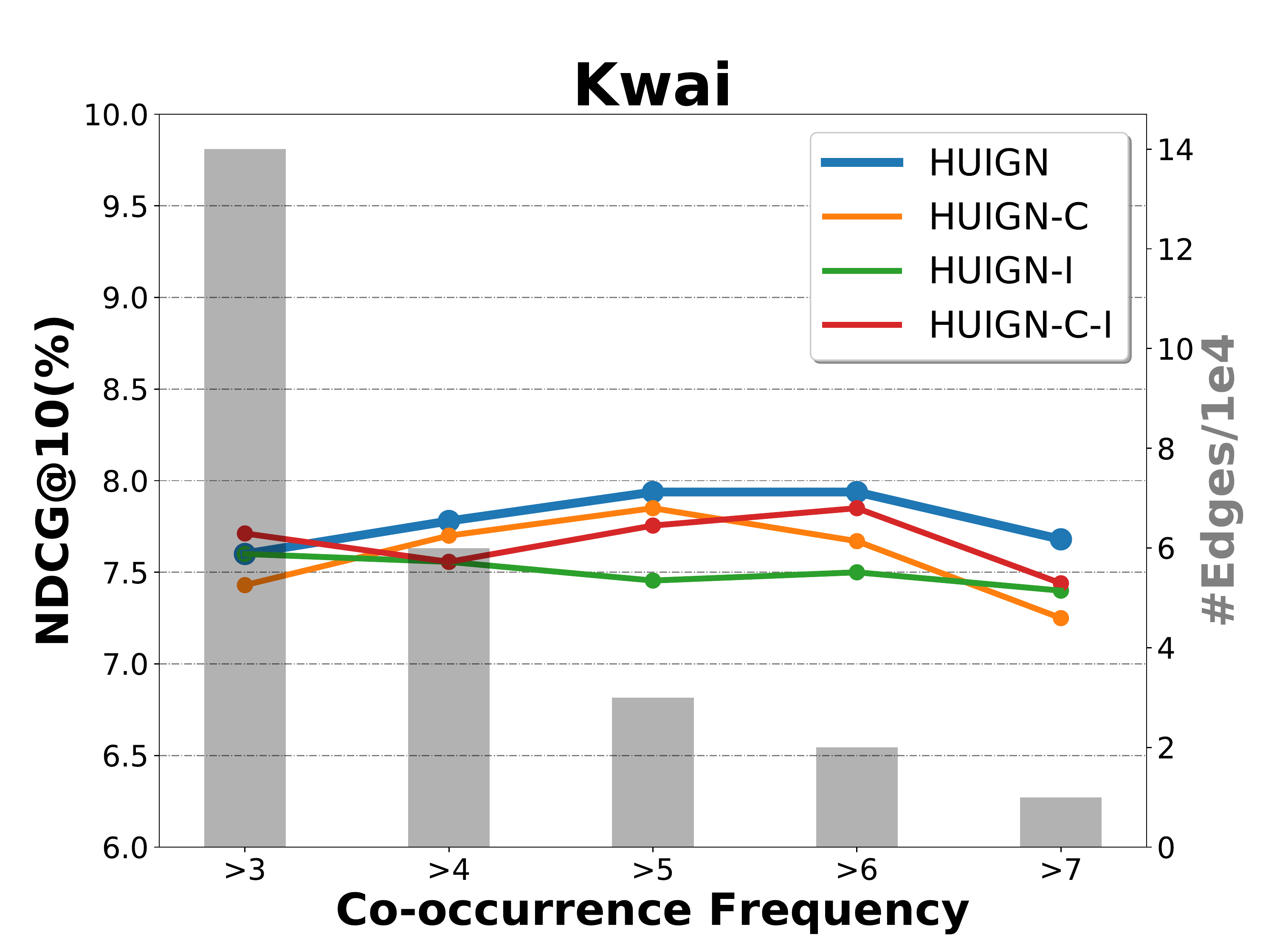}
      \label{fig_visualize_2_3}
    }
    % \vspace{-10pt}
    \caption{Performance in terms of Precision@10, Recall@10 and NDCG@10 w.r.t. different co-interacted frequencies on Movielens and Kwai.~(The histograms indicate the numbers of edges in different frequencies.)}
    \label{fig_3}
    % \vspace{-5pt}
 \end{figure*} 
From the summaries in Table~\ref{table_3}, we capture the following observations:
\begin{itemize}
    \item 
    As a whole, we find that both the number of layers and the number of user intents~(\textit{i.e.} supernodes) in each layer affect the performance of our proposed model. 
    For the range from one to three, increasing the layer could boost the performance on the three datasets in most cases.  
Although the absolute improvements are not significant, we find that the three-layer variants surpass the single-layer ones by 2.5\% and 7.14\% w.r.t. recall@10 on Tiktok and Kwai datasets with the negligible memory and time complexity according to our analysis in Section~\ref{discuss}. Besides, we construct the hierarchical structure to correspond to the multi-level user intents, whose effectiveness is illustrated in Section~\ref{visualization}.
    It is consistent with our suggestion regarding the effect of multi-level user intents in personalized recommendation. 
    \item
    However, compared with the three-layer structures, the user intents consisted of four levels achieves the suboptimal results in terms of the three metrics. 
    We believe that there are two main reasons. 
    One is that the four-layer structure we designed may disagree with the real-world user intents when people watch the video; and the other is that such many levels of user intents could result in overfitting during the training phase.  
    \item
    Comparing the structures with the same numbers of layers, we find that the results of the user intents with $\{32, 8, 4\}$ structure are better than that of the other one.
    Although more parameters are introduced into the model, they do not improve the performance. 
    Jointly considering the case of four-layer structure, it indicates that our model does not benefit from the graph convolutional operations on the deeper or/and wider graph.  
    \item Jointly analyzing Table~\ref{table_2}, we find that even the single-level variant achieves the comparable performance comparing with the strongest baselines. We attribute the improvement to re-express the item with the learned user intents. It also verifies that our proposed method could capture the informative signal by incorporating the item content with the user historical behaviors. 
\end{itemize}

\subsubsection{Effects of Modalities}
In the multimedia recommendation, multi-modal contents are the information sources of the user intents. 
As such, it is necessary to conduct experiments to investigate these critical variables on Movielens and Tiktok, which consists of the visual, acoustic, and textural modalities features. 
In Table~\ref{table_4}, we exhibit the results in different modalities in terms of three metrics to compare with the results of multiple modalities. 
From this table, we have the following findings:
\begin{itemize}
    \item
    The model with only the visual features achieves the best performance compared to models that use only the acoustic or textual information. 
    This observation is consistent with the finding in ~\cite{MMGCN}. 
    It reveals that the most effective content information comes from the visual modality in micro-video recommendation. 
    Besides the fact that the visual contents convey more information than the audio and textual description, we believe that the user pays more attention to the visual cues when she/he browsers the micro-video. 
    In particular, on Movielens, the results in the visual modality exceed the others by a significant margin. 
    We attribute the improvements to the high-dimension feature vectors, which facilitates the user intent modeling.
    \item 
    Unexpectedly, on Movienlens, the model with visual modality features outperforms the one with multi-modal features.  
    Such a phenomenon may be caused by our simplistic approach in combining the multi-modal representations; we leave the investigation of relations among multi-modal user intents for the future work. 
    Even so, compared with the baselines, our proposed model also achieves the best performance, which demonstrates the effectiveness of representations based on user intents. 
\end{itemize}
\subsubsection{Effects of Cross-entropy and Independence loss}
In our model, we introduced the cross-entropy loss and independence loss. 
To demonstrate the effectiveness of these loss functions~(\textit{i.e.} Eq.~\ref{Equ_CrossEntropy} and Eq.~\ref{Equ_Independent}), we performed the experiments on different co-interacted items graphs. 
In particular, according to the different co-interacted frequencies, we built several graphs with different numbers of edges. 
On Movielens and Kwai, we constructed five graphs where the co-interacted frequencies are larger than 3, 4, 5, 6, and 7, respectively. 
The results in terms of three matrices on Movienlens and Kwai are illustrated in Figure~\ref{fig_3}. 
Whereinto, HUIGN-I, HUIGN-C, and HUIGN-C-I denote the variants without independence loss, cross-entropy loss, and both of them, respectively. 
Besides, in the figure, the histograms represent the numbers of edges in different graphs.  
From the figure, we observed that:
\begin{itemize}
    \item 
    Although HUIGN-C outperforms HUIGN-C-I, it is easy to find that their curves \textit{w.r.t.} metrics in the figures follow similar trends in different cases. 
    Except the cross-entropy loss prevents overfitting in the training phase, we believe that the loss benefits the user intents representations. 
    With this loss function, not only the features reflecting the primary user intents are enhanced but also the others are faded during the assignment, which refines the information used to learn the user intents. 
    Thus, the signals of nodes are integrated into the supernodes which are closest to the intents of users who consumed them.
    \item
    With the increasing of co-interacted frequency, the variants without the independence loss are susceptible to the number of edges in the item graph. 
    As shown in the figures, the curves representing the HUIGN-I are more smooth than those of other variants. 
    Hence, we suggest that the independence loss makes the method more robust. 
    The attribute comes from that the independence loss could be used to disentangle the user intents at the same level, which prevents each intent from the noise caused by other intents.  
    \item
    We further compared our proposed model to the variants, which shows the performance of HUIGN is better than the others'. 
    Combined with the above observations, it demonstrates that the advantages of these two loss functions are mutually strengthened. 
    Specifically, with the independence loss, the users' intents can be disentangled into several independent ones, each of which reflects one factor of users' motivations. 
    Based on these intents, each node's assignment vector are easily closing to the one-hot vector, which enhances the user intents representations and benefits the independence.
\end{itemize}
\begin{figure}
	\centering
    \includegraphics[width=0.4\textwidth]{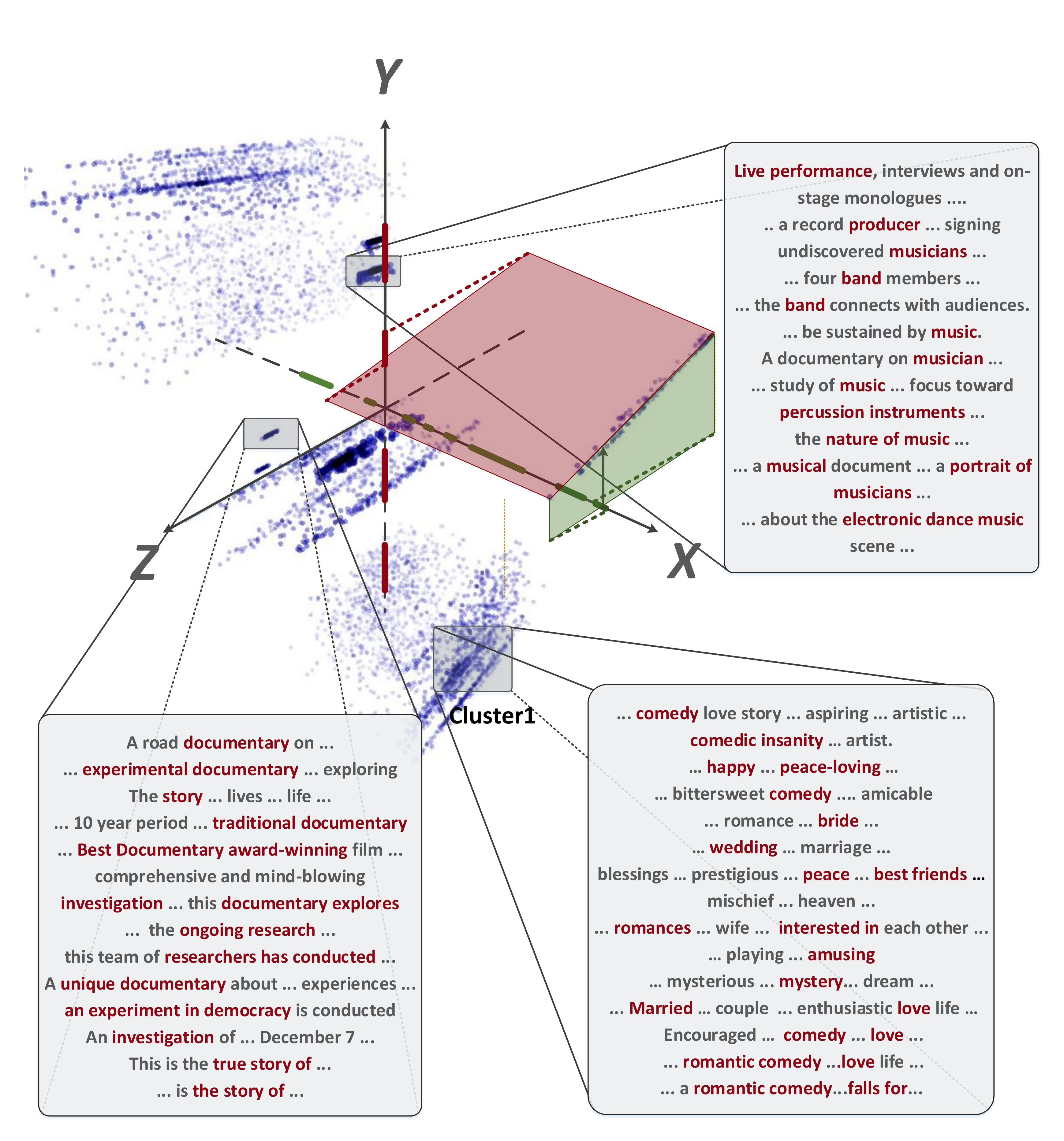}
    \vspace{-10pt}
    \caption{Visualization of items’ representations using the t-SNE dimensionality reduction algorithm in 3D. Whereinto, X, Y, and Z annotate the three axes in the 3-dimensional coordinate, respectively; red and green represent the mapping operations.}
	\label{fig:vis}
\end{figure}
\subsection{Visualization~(RQ3)}
\label{visualization}
In this section, we aim to provide the semantic description for the user intents learned in our proposed model. 
For this purpose, we used the t-Distributed Stochastic Neighbor Embedding~(t-SNE) in 3-dimension to visualize the user intents and introduced some descriptions for them. 
In particular, we performed the t-SNE algorithm on items' characters in the textual modality and leverage the videos' descriptions to offer the semantic information\footnote{Since the raw data in Tiktok and Kwai are not released, we only conducted the experiment on Movienlens.}, as illustrated in Figure~\ref{fig:vis}.  
And, we have the following findings:
\begin{itemize}
    \item
    Observing the distribution of items in the 3D coordinate, we find that most items are clustered into several regions. 
    The observation is consistent with our idea that representing the items according to their characters on several user intents.  
    Further, when we respectively project items to three axes in the coordinate, we could find that they are projected into several separate regions in each axis. 
    More specifically, the distribution on the Y-axis is concentrated in 4 discrete regions and the distribution on the X-axis can be grouped into 8 regions, which follows our designs of multi-level user intents. 
    Although, on the Z-axis, the points do not obviously fall into 32 regions like our setting, we suggest that there is no clear margin between the user intents at the low level.  
    Therefore, we represent each cluster with the items belonging to it. 
    \item 
    To answer the question of what the user intents are, we displayed the videos' descriptions and highlighted the keywords in the figure to semantically exhibit three distinct clusters. 
    From each clusters' keywords, we could infer the main themes of items in this cluster. 
    Taking Cluster1 as an example, we observe that the most frequent words are `\textit{comedy}` and `\textit{love}`.
    It reveals that some users prefer these videos for the comedic elements and love stories in videos. 
    And, for the other displayed clusters, users' interests are towards `\textit{music}` and `\textit{documentary}`. 
    By jointly analyzing with the above finding, we are able to provide the semantic representation for each user intent with the keywords in videos' descriptions. 
\end{itemize}
\section{Conclusion and Future Work}
In this work, we proposed to model user intents as well as their structure for user and item representation learning in the multimedia recommender system. 
For this purpose, we developed a new framework based on graph convolutional networks, which explicitly models a hierarchical graph to represent the multi-level user intents. 
By performing the information aggregation operations on the constructed co-interacted item graph, the model learns the user intent implied in the historical interaction records and builds the relationship between the finer-grained and coarser-grained user intents. 
Ultimately, the user and item characteristics based on the user intents provide us with the explainable and high-quality recommendations. 

To the best of my knowledge, this is the first attempt to explicitly model the user intent without any external information, such as user social information and demographics, in the recommender system. 
In future, there are still many extensions that we need to continue to explore. 
For instance, we would further investigate the relationship of user intents among the multiple modalities. 
Although the user preference is dependent on different modalities, we believe the consistency probably emerges at the general level. 
Besides, we extend our work by introducing external users' social signals and items' knowledge information, which aims to learn an explainable connection among the user intents, social relations, and item knowledge. 
\bibliography{BIB/IEEEabrv, reference}

\end{document}